\documentclass[%
 reprint,
superscriptaddress,
 amsmath,amssymb,
 aps,
 prl
]{revtex4-1}
\usepackage{upgreek,color}
\usepackage{graphicx}
\usepackage{dcolumn}
\usepackage{bm}

\begin{document}

\preprint{APS/123-QED}

\title{Anomalous blue-shift of terahertz whispering-gallery\\ modes via dielectric and metallic tuning}

\author{Dominik Walter Vogt}
\affiliation{Department of Physics, The University of Auckland, Auckland 1010, New Zealand}%
\affiliation{The Dodd-Walls Centre for Photonic and Quantum Technologies, New Zealand}
 \email{d.vogt@auckland.ac.nz}
\author{Angus Harvey Jones}%
\affiliation{Department of Physics, The University of Auckland, Auckland 1010, New Zealand}%
\affiliation{The Dodd-Walls Centre for Photonic and Quantum Technologies, New Zealand}
\author{Harald G. L. Schwefel}%
\affiliation{Department of Physics, University of Otago, 730 Cumberland Street, Dunedin 9016, New Zealand}
\affiliation{The Dodd-Walls Centre for Photonic and Quantum Technologies, New Zealand}
\author{Rainer Leonhardt}%
\affiliation{Department of Physics, The University of Auckland, Auckland 1010, New Zealand}%
\affiliation{The Dodd-Walls Centre for Photonic and Quantum Technologies, New Zealand}






\date{\today}

\begin{abstract}
The vast majority of resonant systems show a red-shift for the resonance frequency when a perturbation, e.g. losses, are introduced to the system. In contrast, here we report for the first time the experimental demonstration of both red- and anomalous blue-shifting of whispering-gallery modes (WGMs) using dielectric and metallic substrates. The maximum blue-shift is more than three times as large as the expected red-shift, proving that the anomalous blue-shift is more than a peculiar curiosity. The experiments are performed in the terahertz (THz) frequency range with coherent continuous-wave spectroscopy. The results establish dielectric and metallic tuning as a novel, and viable approach to tune high quality (high-Q) WGMs, and provide valuable insights into the anomalous blue-shift of WGM cavity systems. The tuning capabilities for these compact monolithic resonators is of significant interest for fundamental science and technological applications alike.
\end{abstract}

\maketitle



Whispering-gallery mode resonators (WGMRs) continue to attract tremendous interest across various fields and technologies due to their compact size and ultra high-Q factors. Typical applications include sensing, quantum electrodynamics, optical communication, and frequency up-conversion to name but a few \cite{strekalov2016nonlinear,heebner2008optical,matsko2006optical,ilchenko2006optical}. However, the fixed resonance frequencies determined by the geometry and optical properties of the dielectric resonator limit their functionality for many applications, in particular for light-matter interaction.

Typically, tuning of the WGM resonance frequency is achieved by exploiting the temperature dependent refractive index of the dielectric, or mechanically deforming the resonator with strain - both methods are directly changing the optical path length of the WGM and therefore the resonance frequency \cite{ioppolo2007pressure,ilchenko1998strain,von2001frequency,pollinger2009ultrahigh,vogt2018thermal}. More recently, a theoretical investigation predicted tuning of the WGM resonance frequencies by bringing a dielectric substrate (modelled as half-plane) close to the evanescent field of the WGM. In particular, besides a well-known red-shift, the theory predicts an anomalous blue-shift of the WGM \cite{foreman2016dielectric,ruesink2015perturbing}.

The blue-shift is particularly surprising, as any increase of the average refractive index seen by the evanescent field -- i.e. an increasing in the optical path length -- of the WGM is expected to red-shift the resonance frequency. In contrast, an anomalous blue-shift of the WGM has to be related to a decrease in the optical path length, e.g, by forcing the WGM further into the resonator \cite{grosjean2012extraordinary}. This however can be achieved under certain circumstances via the boundary conditions at the surface of the dielectric substrate. The phase and the amplitude of the reflected radiation at the boundary can either lead to a phase shift of the resulting WGM or can shift the mode centre slightly toward the centre of the resonator. When the refractive index of the dielectric substrate is large enough, the latter dominates the effect of a larger average refractive index seen by the evanescent field, ultimately leading to an anomalous blue-shift of the WGM. Consequently, we expect the strongest blue-shift not for a dielectric but for a metallic substrate. This is most obvious when the metallic substrate touches the resonator, as there has to be a node of the electric field at the point where the resonator touches the metallic substrate -- shifting the center of the WGM towards the center of the resonator.  


Here, we present the first experimental observations of the predicted blue-shift in the terahertz (THz) frequency range using dielectric substrates. Furthermore, we demonstrate unprecedented results achieving an anomalous blue-shift of THz WGMs using metallic substrates. While the authors of reference \cite{foreman2016dielectric} only show the maximum frequency-shift as a function of the refractive index, we also measure the frequency-shift as a function of the distance between the dielectric/metallic substrates and the WGMR. The results are expected to be applicable to any other frequency range, and provide a viable approach to tune WGM resonances, beneficial for many applications and technologies based on WGMRs. Finally, our results establish novel insights into the anomalous blue-shift of WGM cavity systems.

\begin{figure}[t!]
\centering
\includegraphics[width=\linewidth]{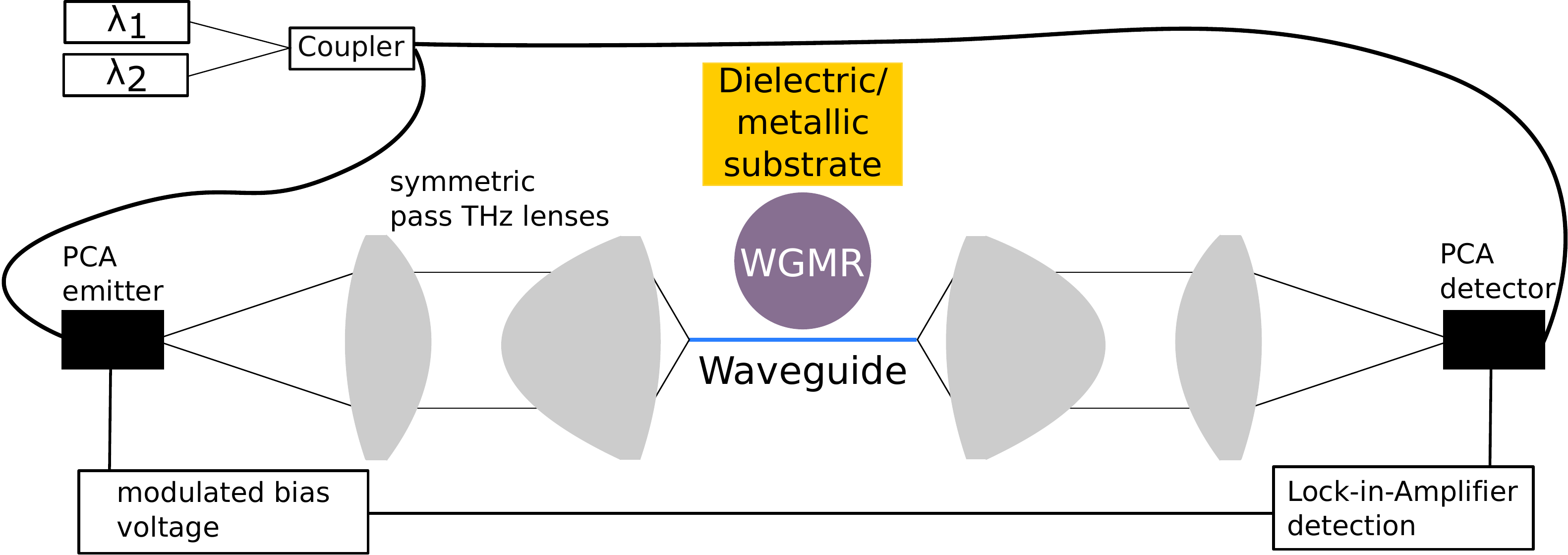}
\caption{Scheme of the experimental setup. The THz radiation is collimated and focused by specially designed symmetric-pass UHWMPE lenses \cite{lo2008aspheric}. The spectrometer has an effective frequency resolution of about 4\,MHz.}
\label{fig:1}
\end{figure}

We investigate the frequency tuning of the THz WGMs using a variety of dielectric as well as metallic substrates both experimentally and numerically. A scheme of the experimental setup based on coherent continuous wave THz spectroscopy is shown in Fig. \ref{fig:1}. The THz radiation is generated and detected using commercially available fiber-coupled photo-conductive antennas (PCAs) (Toptica) \cite{deninger20152}. The polarization is chosen to be out of the plane of Fig. \ref{fig:1}, thus only transverse electric (TE) WGMs are excited. TE WGMs are expected to show a much larger blue-shift than transverse magnetic (TM) WGMs \cite{foreman2016dielectric}. The experimental setup and the data analysis based on Hilbert transformation is explained in detail in previous publications \cite{vogt2018coherent,vogt2018ultra,vogt2017high}. The WGMR under study is a 4\,mm diameter ($r_0=2\,\textrm{mm}$) high-resistivity float-zone grown silicon (HRFZ-Si) sphere with a material refractive index of $n=3.416$ at 0.6\,THz \cite{vogt2018thermal}.

The WGMs are excited using the evanescent field of a 200\,$\upmu$m diameter single-mode air-silica step-index waveguide. The effective refractive index of the fundamental mode of the waveguide is about 1.3 at 0.6\,THz \cite{vogt2018ultra}. Consequently, only higher order radial modes like the ${\textrm{TE}}_{36,12,0}$ WGM at 618.9\,GHz, with $N=36$ wavelengths along the circumference, and 12 nodes in radial direction of the WGMR, are excited. The WGMR is mounted on a 3D computer controlled stage, which allows an adjustment of the waveguide-WGMR distance to achieve the desired coupling.

The measurement procedure is as follows: initially, the THz WGM is slightly over coupled, and the waveguide-WGMR distance is kept constant throughout further measurements. First, the dielectric/metallic substrate, mounted on a translation stage, is brought into contact with the WGMR, which leads to a minute reversible movement of the WGMR. Subsequently, the dielectric/metallic substrate is withdrawn until the WGMR is observed to be in its original position -- this is defined as zero distance $d$ between the WGMR and the dielectric/metallic substrates. The position of the WGMR is monitored with 20x magnification microscope connected to a CCD camera, allowing identification of movements of the WGMR smaller than 0.5\,$\upmu$m. Next, the induced frequency shift is measured for a series of distances $d$ in the range from 10\,$\upmu$m to 200\,$\upmu$m. At each distance the shifted resonance frequency is obtained from a fit to the measurements. The fit is based on an analytic model describing the evanescent coupling of WGMs \cite{gorodetsky1999optical}. The frequency shifts are calculated in respect to the resonance frequency of the over coupled WGM and plotted as a function of the distance $d$. The obtained curve shows an exponential behaviour as expected from previous analytic models describing the frequency-shift of WGMs due to the presence of a coupling-prism \cite{demchenko2017optimisation}. 

Next, the intercept at $d=0$ of an exponential fit to the experimental data is extracted to calculate the frequency shift compared to the intrinsic WGM resonance frequency $f_0$. Due to the presence of the air-silica step-index waveguide, $f_0$ is slightly shifted compared to the over coupled WGM resonance frequency (< 5\,MHz). The data points at 10\,$\upmu$m and 30\,$\upmu$m have a low signal to noise ratio (SNR) and are therefore excluded from the exponential fit. The latter is only applicable to dielectric substrates, since metallic substrates show considerably better SNR, as discussed below. Finally, the entire process is repeated at least three times for each dielectric/metallic substrate. In total five dielectric substrates with a range of refractive indices are investigated, namely: Polytetrafluoroethylene (PTFE) ($n=1.4$), ultra-high-molecular-weight polyethylene (UHWMPE) ($n=1.52$), quartz ($n=1.96$), HRFZ-Si ($n=3.416$), and birefringent rutile (ordinary ${\textrm{n}}_{\textrm{o}}=9.35$, extraordinary ${\textrm{n}}_{\textrm{e}}=12.59$) \cite{jin2006terahertz,lo2008aspheric,naftaly2007terahertz,vogt2018thermal,jordens2009terahertz}. The minute birefringence of quartz has been neglected. Furthermore, we investigate the following metallic substrates: copper, aluminum and gold. In the THz frequency range the metallic substrates can effectively be treated as perfect electric conductors, and consequently very similar frequency-shifts are expected for all metallic substrates \cite{maier2006terahertz}.      


The experimental results are qualitatively supported with finite-element method (FEM) simulations performed with Comsol Multiphysics\textsuperscript{\textregistered}. Due to computational requirements, the simulations are restricted to 2D with a cylindrical HRFZ-Si 4\,mm diameter WGMR and a slab air-silica step index waveguide. Nevertheless, the simulations provide valuable insights into the tuning of THz WGMs using dielectric and metallic substrates.

\begin{figure}[t!]
\centering
\includegraphics[width=\linewidth]{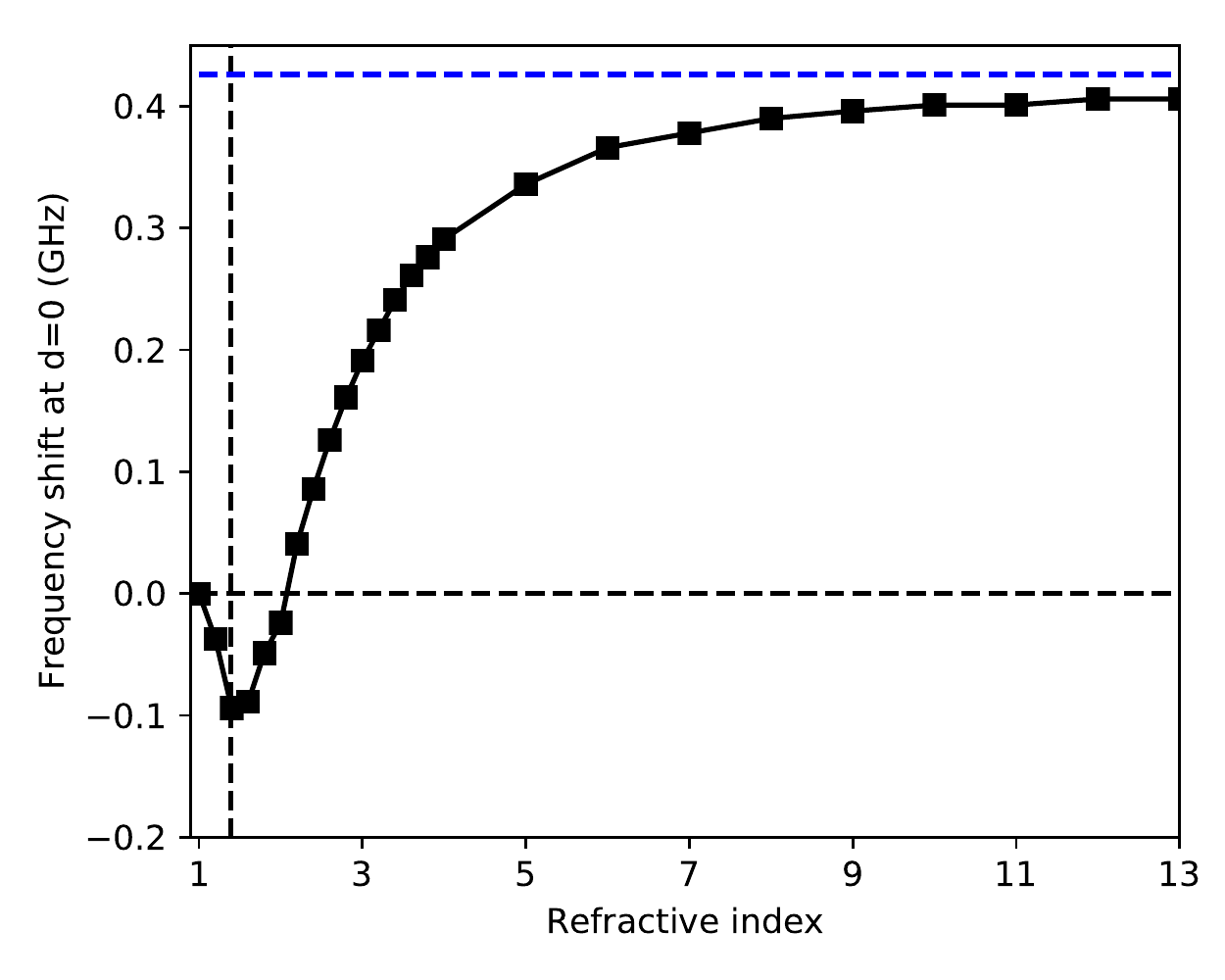}
\caption{Simulated frequency shift for the ${\textrm{TE}}_{36,12,0}$ WGM at $d=0$. The black dashed horizontal line shows zero frequency shift to guide the eye. The vertical dashed lines show the corresponding phase refractive index of the WGM. The blue dashed line indicates the simulated blue-shift with a perfect electric conductor.}
\label{fig:2}
\end{figure}


The numerical results presented in Fig. \ref{fig:2} demonstrate the simulated frequency shift as a function of the material refractive index of the dielectric substrate at $d=0$. The simulations are performed for the 12th higher order radial WGM at 613.4\,GHz, corresponding to the WGM used in the experiment. As expected, no frequency shift is observed for a refractive index of 1. First, with increasing material refractive index, an increasing red-shift is observed. Maximum red-shift occurs slightly above the phase refractive index ${n}_{\textrm{ph}|{r}_{0}}$ of the WGM at the surface of the resonator. This observation is in accordance with the analytic model, where the maximum red-shift is expected at a material refractive index slightly above the effective refractive index of the WGM. The phase refractive index ${n}_{\textrm{ph}|{r}_{0}}$ is simply calculated with ${n}_{\textrm{ph}|{r}_{0}} = N\textrm{c}/2\uppi {r}_{0} f_0$ - the condition of constructive interference of the WGM after propagation along the circumference of the resonator; with c the speed of light. The corresponding phase refractive index of 1.4 is indicated in Fig. \ref{fig:2} with the vertical dashed line. The investigated THz WGMs are described using a localized phase refractive index as a function of the radial distance rather than an effective refractive index. An effective refractive index is inappropriate due to the large mode volume in relation to the size of the spherical resonator \cite{vogt2018ultra}. The phase refractive index at the surface of the WGMR has been chosen as this is a point of strong interaction with the dielectric substrate. 

Next, with further increasing material refractive index, the red-shift is diminished until a blue-shift is observed. In accordance with the theory, the frequency shift as a function of the refractive index flattens with increasing material refractive index. As expected, the maximum blue-shift with the dielectric substrates asymptotically approaches the blue-shift obtained for a perfect electric conductor in contact with the WGMR (horizontal dashed blue line in Fig. \ref{fig:2}). The special case of a metallic tuning is discussed in more detail below.


\begin{figure}[t!]
\centering
\includegraphics[width=\linewidth]{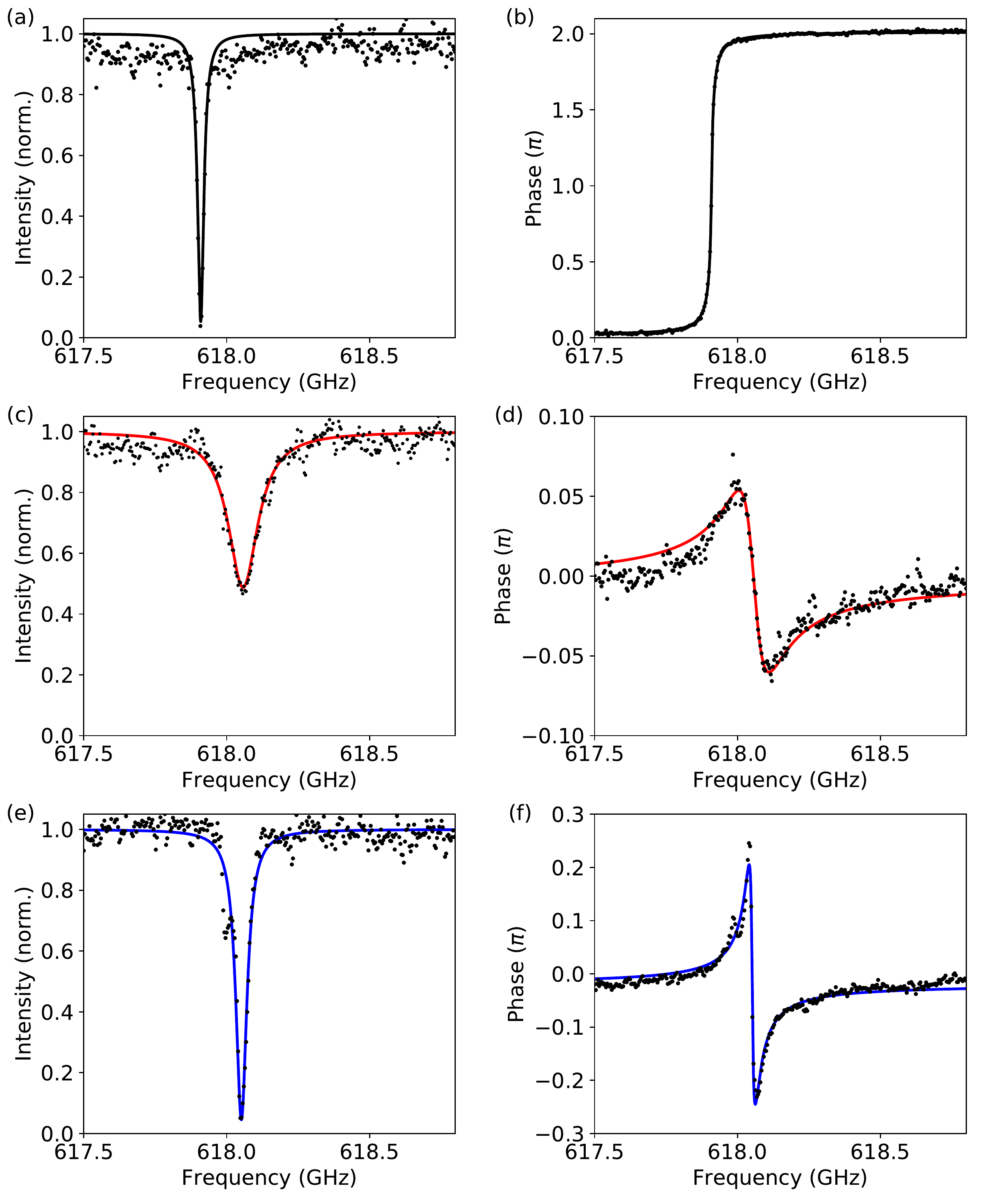}
\caption{Measured intensity and phase profiles (black dots) of the ${\textrm{TE}}_{36,12,0}$ WGM with (a,b) and without (c,d) rutile (extraordinary axis) dielectric substrate at a distance $d=30\,\upmu$m, respectively. The black and red solid lines show the fitted analytic model with air and rutile, respectively. (b) shows the typical 2$\uppi$ phase transition for an over coupled WGM, while (d) shows the phase profile associated with an under coupled WGM. Sub-figures (e) and (f) show the measured intensity and phase profile of the same THz WGM but with and without an aluminum substrate at $d=30\,\upmu$m, respectively. The corresponding fit is shown with blue solid lines. The presented data is averaged from three independent measurements.}
\label{fig:3}
\end{figure}

In the following the experimental results for the dielectric and metallic tuning of the THz WGMs are presented. Figure \ref{fig:3} shows exemplary intensity and phase profiles (black dots) of the 12th higher order radial mode at 617.9\,GHz without (a, b) and with (c, d) rutile (extraordinary axis), respectively. 
An excellent agreement is observed for the fit with the complex analytic model. The blue-shift induced by the rutile at a distance $d=30\upmu$m is clearly visible by comparing of Fig. \ref{fig:3}(a) and Fig. \ref{fig:3}(c). The obvious mode broadening is due to additional radiation losses caused by the dielectric substrate. Also, because of the additional losses, the WGM undergoes a transition from over coupling to under coupling, as can be seen from the phase profiles [see Figs. \ref{fig:3}(b) and \ref{fig:3}(d)] \cite{vogt2017terahertz}. 

For comparison, Figs. \ref{fig:3}(e) and \ref{fig:3}(f) show the corresponding intensity and phase profiles, respectively, for an aluminum substrate. At the same distance of $d=30\upmu$m the linewidth and SNR is significantly improved. We attribute the sharpness of the resonance with the metallic substrate to the fact that there are no losses related to the transmission into the material as there are for the dielectric substrates. However, the observed small broadening of the linewidth is expected as the metal surface will reflect some of the evanescent field that will not constructively interfere with the WGM. 


\begin{figure}[t!]
\centering
\includegraphics[width=\linewidth]{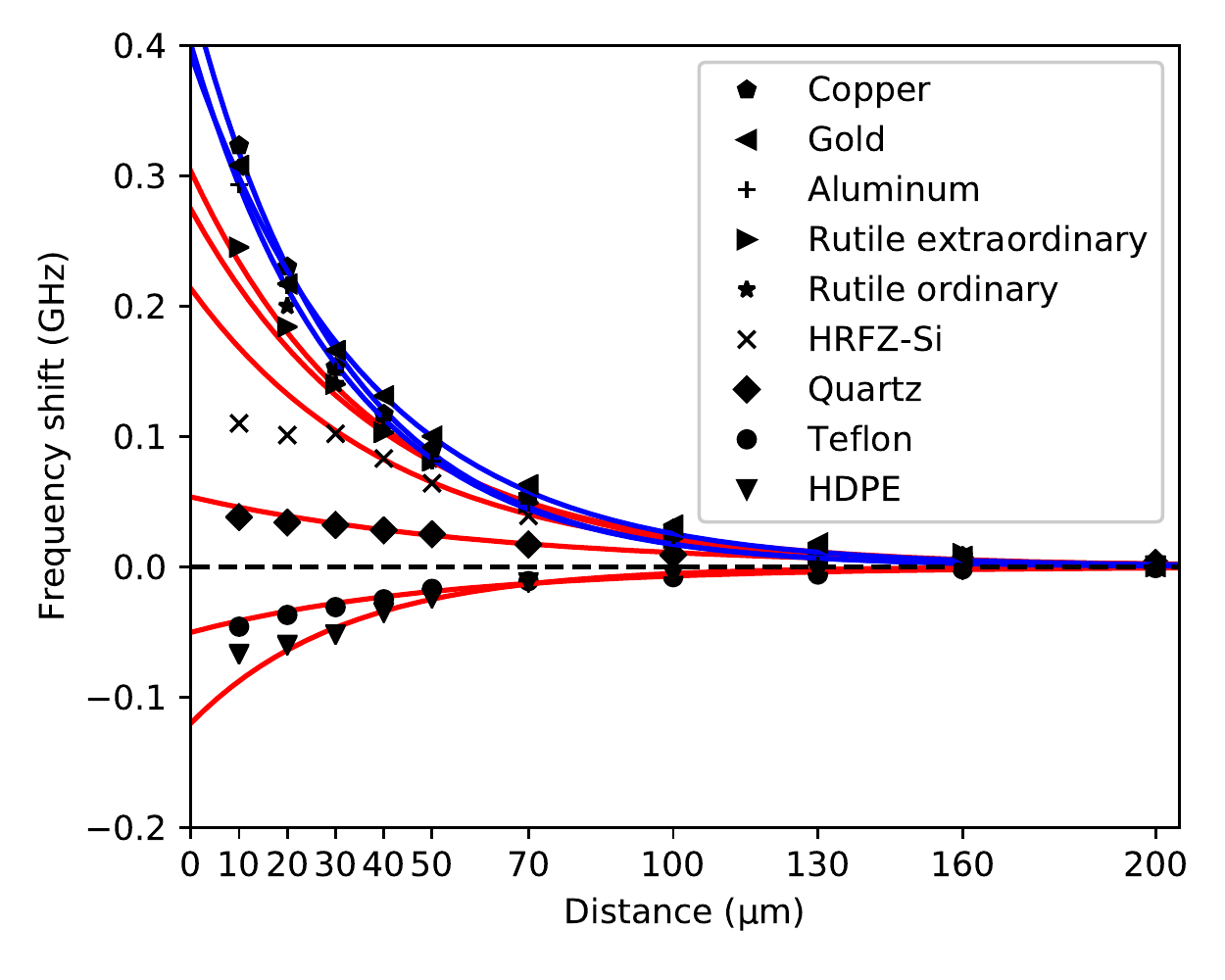}
\caption{Measured frequency shift (symbol) as a function of $d$ exemplary for a single measurement. The corresponding exponential fit is shown with solid red lines for dielectric substrates and blue solid lines for metallic substrates. The horizontal black dashed line indicates zero frequency shift to guide the eye.}
\label{fig:4}
\end{figure}

The measured frequency shifts (symbols) as a function of the distance $d$ and the corresponding exponential fits (lines) are plotted in Fig. \ref{fig:4} for all measured dielectric and metallic substrates. The graph beautifully shows both red- and blue-shifts depending on the material refractive index of the dielectric substrates. However, all three metallic substrates demonstrate the strongest blue-shift and show a very similar exponential behavior.



\begin{figure}[t!]
\centering
\includegraphics[width=\linewidth]{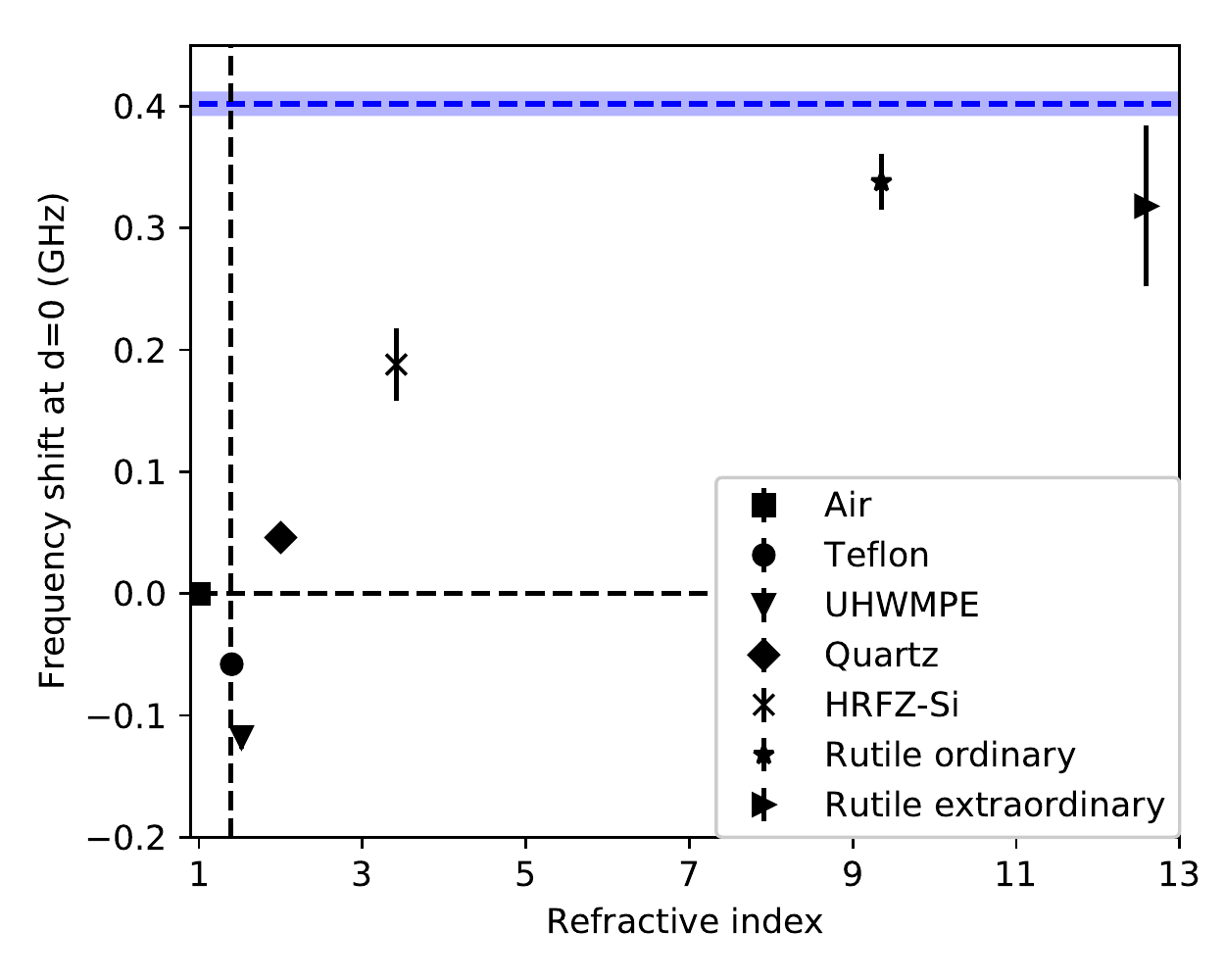}
\caption{Measured frequency shift (symbols) as function of the material refractive index for the dielectric substrates. The WGM phase refractive index and zero frequency shift are indicated with vertical and horizontal dashed lines to guide the eye. The errobars show the corresponding standard deviations. The blue dashed line indicates the average blue-shift obtained with the metallic substrates and the corresponding standard deviation (light blue shaded area).}
\label{fig:5}
\end{figure}

In the following, we extract the frequency shifts at $d=0\upmu$m (intercepts in Fig. \ref{fig:4}) from the exponential fits of the dielectric substrates and plot them in Fig. \ref{fig:5} over the corresponding material refractive indices. The curve shows an excellent qualitative agreement with the previously presented simulations. Please note that a quantitative agreement is not expected as the simulations are only performed in 2D. Maximum red-shift is observed at a refractive index ${n}_{\textrm{substrate}}$ slightly above ${n}_{\textrm{ph}|{r}_{0}}$. In physical terms, the increase in red-shift for ${n}_{\textrm{substrate}} < {n}_{\textrm{ph}|{r}_{0}}$ can be attributed to the reflected radiation enhancing the average refractive index seen by the evanescent field. For ${n}_{\textrm{substrate}} > {n}_{\textrm{ph}|{r}_{0}}$ there will be an additional phase shift of $\uppi$, and therefore the interference between the reflected radiation and the WGM will now lead to an anomalous blue-shift as theoretically predicted for dielectric substrates. The measured frequency shifts for the dielectric substrates are as follows: -58$\pm$5\,MHz (teflon, $n=1.4$), -118$\pm$9\,MHz (UHWMPE, $n=1.52$), 46$\pm$1\,MHz (quartz, $n=2.0$), 188$\pm$30\,MHz (HRFZ-Si, $n=3.416$), 338$\pm$23\,MHz (rutile ordinary, ${\textrm{n}}_{\textrm{o}}=9.35$), and 318$\pm$66\,MHz (rutile extraordinary, ${\textrm{n}}_{\textrm{e}}=12.55$). The measured blue-shift for the dielectric tuning is about three-times as large as the maximum observed red-shift.  

The measured blue-shifts with the metallic substrates are: 408$\pm$7\,MHz, 388$\pm$5\,MHz, and 411$\pm$7\,MHz for copper, aluminum, and gold, respectively. The metallic substrates show consistently the strongest blue-shifts with an average value of 402\,MHz$\pm$10\,MHz -- the maximum value asymptotically approached by the dielectric tuning (see Fig. \ref{fig:5}). We attribute the maximum blue-shift to the boundary conditions introduced by the metallic substrate: since the electric field of the WGMs has to be zero at the point where the metallic substrate is in contact with the WGMR, the electric field is shifted towards the center of the WGMR and therefore reducing the optical path length -- ultimately leading to the observed anomalous blue-shift.            




In conclusion, the presented experimental results successfully verify the theoretically predicted anomalous blue-shift of WGMs using dielectric substrates. Furthermore, the unprecedented experiments with metallic substrates provide novel insights into the anomalous blue-shift of WGM cavity systems. In particular, the anomalous blue-shift of dielectric substrates is found to asymptotically approach the blue-shift obtained with metallic substrates. Compared to dielectric tuning, metallic tuning offers the largest frequency shift while retaining high-Q WGMs - rendering metallic tuning a particularly viable approach for numerous applications and technologies where tuning of WGMs is essential.







%



\end{document}